# MAC Performance and Algorithmic Optimization in Matrix Multiplication Workloads

Daniel Negasa
Department of Computer Science and Engineering
American University of Ras al Khaimah
Ras al Khaimah, United Arab Emirates
2023006224@aurak.ac.ae

Nada Osama Mohamed
Department of Computer Science and Engineering
American University of Ras al Khaimah
Ras al Khaimah, United Arab Emirates
nada.mohamed@aurak.ac.ae

Arfan Ghani
Department of Computer Science and Engineering
American University of Ras al Khaimah
Ras al Khaimah, United Arab Emirates
arfan.ghani@aurak.ac.ae
ORCID:0000-0002-1362-3720

***Abstract*—Matrix multiplication is a fundamental computational kernel underlying a wide range of real-world applications, including machine learning, scientific computing, signal processing, and computer graphics. Its performance directly impacts the efficiency, scalability, and energy consumption of modern computing systems. This paper presents a comparative analysis of several matrix multiplication algorithms implemented in software and examined in the context of their hardware execution characteristics. Naive, NumPy, Strassen, and Winograd algorithms are evaluated based on execution time, user time, and CPU time across increasing matrix sizes. The performance metrics reveal computational bottlenecks and highlight the benefits of algorithmic optimizations. Furthermore, the study investigates the mathematical operations underlying each algorithm and analyzes how matrix dimensions influence MAC (Multiply-Accumulate) behavior and overall computational efficiency in the hardware domain. The results provide a performance benchmark and contribute to understanding how algorithmic choices interact with modern computing architectures for applications in computer architecture, data science, and real-time embedded systems.**

## I. INTRODUCTION

Matrix multiplication is a foundational operation in scientific computing, deep learning, and computer graphics, and its performance directly affects the efficiency of entire systems. As applications scale in size and complexity, such as large-scale neural networks and graphics rendering, efficient matrix multiplication becomes increasingly critical. Over the years, many algorithms have been proposed to accelerate matrix multiplication. The simplest 'naïve' algorithm is straightforward but scales cubically with matrix size; for large matrices, performance becomes a major bottleneck. More advanced libraries, such as high-performance dense linear algebra libraries, implement optimized routines that exploit cache hierarchies, blocking, SIMD/vector instructions, and tailored micro-kernels to improve throughput [1][2].

Beyond classical routines, asymptotically faster algorithms such as Strassen's algorithm and Winograd's algorithm reduce the number of arithmetic operations compared to naive multiplication. Strassen's algorithm reduces computational complexity below cubic time by a divide-and-conquer approach, while Winograd optimizes the mix of multiplications and additions. These have long promised faster performance for sufficiently large matrices [3][4].

However, despite their theoretical appeal, in practice, their benefits are often limited. Factors such as memory hierarchy inefficiencies, recursion overhead, data layout, and communication costs often degrade performance, especially on real hardware architectures with multi-level caches and complex memory latency behavior [2][5]. This has led many high-performance linear algebra libraries to favour carefully tuned blocked algorithms rather than algorithmically faster methods, because practical performance depends heavily on hardware-specific optimizations, not solely on arithmetic complexity [1][5].

Moreover, while many studies have evaluated matrix multiplication performance on particular hardware such as CPUs and GPUs, the landscape is changing rapidly: modern architectures, multicore CPUs, SIMD/AVX, many-core GPUs, and specialized accelerators introduce new variables, cache architecture, vector units, memory bandwidth, parallelism, and precision trade-offs that strongly influence performance. Recent work demonstrates significant gains by optimizing general matrix-matrix multiplication (GEMM) kernels for the cache hierarchy and by adapting micro-kernels to operand shapes and hardware targets [6]. Others compare CPU vs GPU implementations, showing dramatic speedups on GPUs for large matrices, thereby underscoring the impact of hardware acceleration for data-parallel workloads [7][8].

Despite advances, there remains a gap where only a few comprehensive studies systematically evaluate classic and optimized software algorithms across a range of matrix sizes and shapes, while also analyzing how matrix dimensions influence low-level hardware behavior such as MAC throughput, cache effects, SIMD utilization, and comparing these results against contemporary hardware-accelerated baselines. There is limited literature offering a holistic benchmark that connects software algorithm choices with their hardware execution behavior under varying real-world conditions.

To fill this gap, this work undertakes a comparative evaluation of multiple matrix multiplication algorithms across a variety of matrix sizes and operand shapes. We measure execution time, CPU time, and user time, aiming to capture baseline performance characteristics and processor behavior, not just high-level runtime. In addition, we analyze how the mathematical structure and dimension of matrices influence multiply–accumulate (MAC) behavior and overall computational efficiency. In this paper, we implement selected algorithms on hardware to evaluate hardware-accelerated performance and explore fine-grained parallelism.

FPGAs allow customized data paths, highly parallel MAC units, and pipeline architectures that can significantly accelerate matrix operations while providing energy-efficient computation compared to general- purpose CPUs and GPUs. FPGA implementations enable us to study how algorithmic optimizations translate to specialized hardware, revealing interactions between software design and hardware-level execution that cannot be captured on standard processors alone. This study aims to produce a set of performance benchmarks that: (1) reveal algorithmic bottlenecks and benefits; (2) elaborate software-hardware interactions across CPUs and FPGAs; and (3) provide guidance for future algorithm design and system- level optimizations in high-performance matrix computations. There are several applications in healthcare where hardware acceleration could significantly benefit from these insights [9-11].

## II. METHODOLOGY

### A. Experimental Setup

In this section, we evaluate and compare the performance of four matrix multiplication algorithms, namely, Naive, NumPy, Strassen, and Winograd, using a controlled software environment. All experiments were conducted on a Windows machine using Python 3, with NumPy as the optimized library baseline and Matplotlib for visualization. Square matrices of sizes ranging from 64×64 to 512×512 were tested, with each experiment repeated three times to reduce measurement noise. Execution time was measured using both user time and CPU time, capturing wall-clock duration including system delays, and actual CPU processing time excluding idle periods, respectively. By maintaining a consistent runtime environment across all implementations, the study isolates algorithmic performance differences and highlights the impact of optimization techniques such as recursive divide-and-conquer in Strassen and algebraic reduction in Winograd. Performance metrics, including execution time and speedup relative to the Naive baseline, were computed and analyzed to provide insights into scalability, efficiency, and computational bottlenecks.

For consistency, all implementations are executed in the same runtime environment, where user time captures wall-clock time and includes I/O or system delays, and CPU time reflects actual CPU processing time, excluding idle or I/O wait. NumPy serves as the baseline optimized library. Strassen's algorithm employs recursive divide-and-conquer, while Winograd's algorithm reduces the number of multiplications using algebraic transformations.

Execution time is plotted for each method, and speedup over the Naive baseline is calculated with equation 1:

$$Speedup_{method} = \frac{T_{naive}}{T_{method}} \quad (1)$$

### B. Algorithmic Performances

#### *i.* ***NumPy***

NumPy relies on optimized C and BLAS routines. Its practical performance exceeds most hand-coded approaches due to better cache locality and SIMD support.

#### *ii.* ***Strassen's Algorithm***

Strassen's method recursively divides matrices into quadrants and combines 7 matrix products instead of 8. Its theoretical complexity is approximately $O(n^{2.81})$, but the overhead and recursion depth must be managed carefully.

#### *iii.* ***Winograd's Algorithm***

Winograd's variant reduces the number of multiplications by introducing intermediate sums and products. It is faster for moderate sizes but suffers from more complex indexing.

## III. RESULTS AND DISCUSSION

### A. Experiments

This section presents a comprehensive performance comparison of various matrix multiplication algorithms, including Naive, Strassen, Winograd, and NumPy. We examine their execution time characteristics across increasing matrix sizes and value ranges. Both user time and CPU time are analyzed to understand computational cost and resource usage. The following figures highlight key insights, trends, and efficiency benchmarks observed during the experiments.

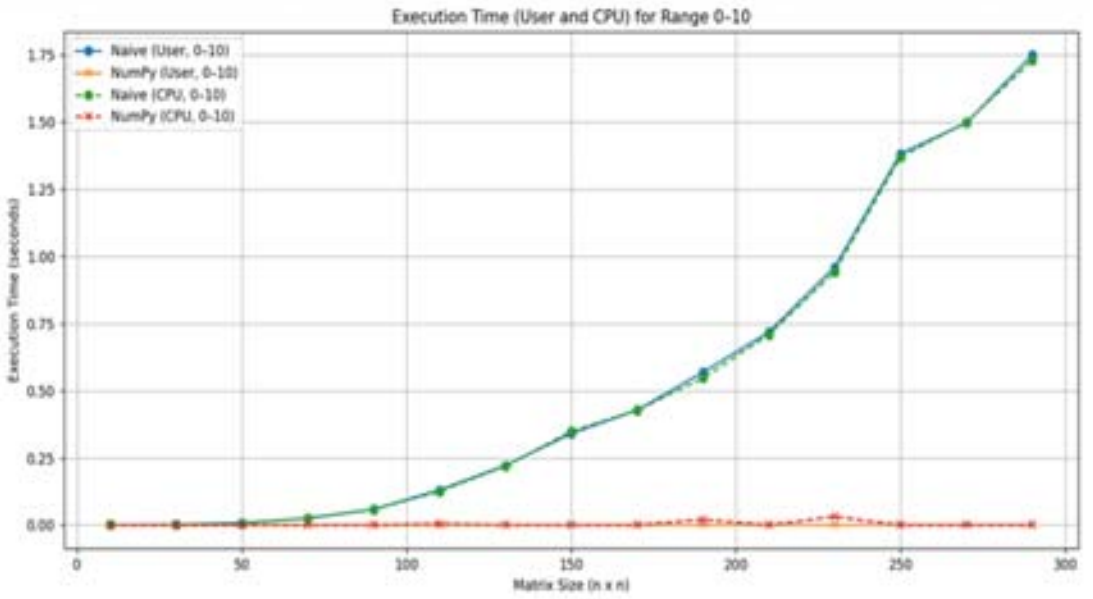

**Fig. 1:** Execution Time (User and CPU) of Naive and NumPy Matrix Multiplication Algorithms for Value Range 0–10.

Figure 1 shows the execution time (both user and CPU) of Naive and NumPy implementations across increasing matrix sizes for inputs ranging between 0 and 10. It clearly demonstrates that NumPy significantly outperforms Naive in terms of both user and CPU time, remaining nearly constant while the Naive method exhibits a quadratic or higher growth.

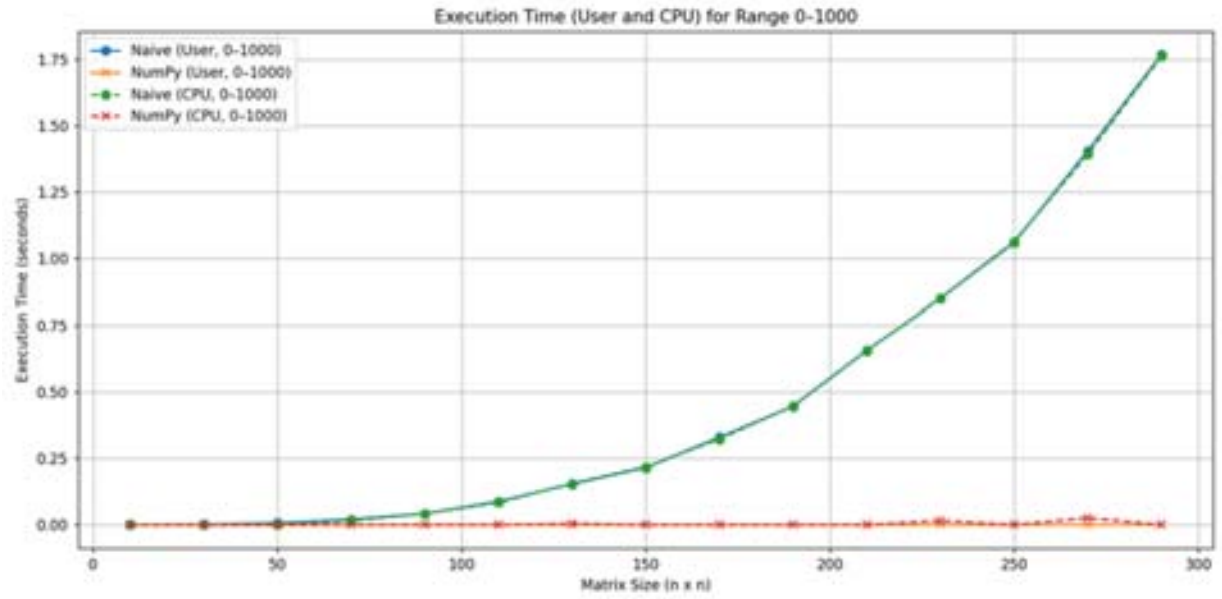

**Fig. 2:** Execution Time (User and CPU) of Naive and NumPy Matrix Multiplication Algorithms for Value Range 0–1000.

Figure 2 repeats the above experiment with input values in the range 0 to 1000. The overall trends remain similar, but the higher range increases numerical computation cost slightly. Still, NumPy remains consistently fast while Naive shows rapidly increasing execution time.

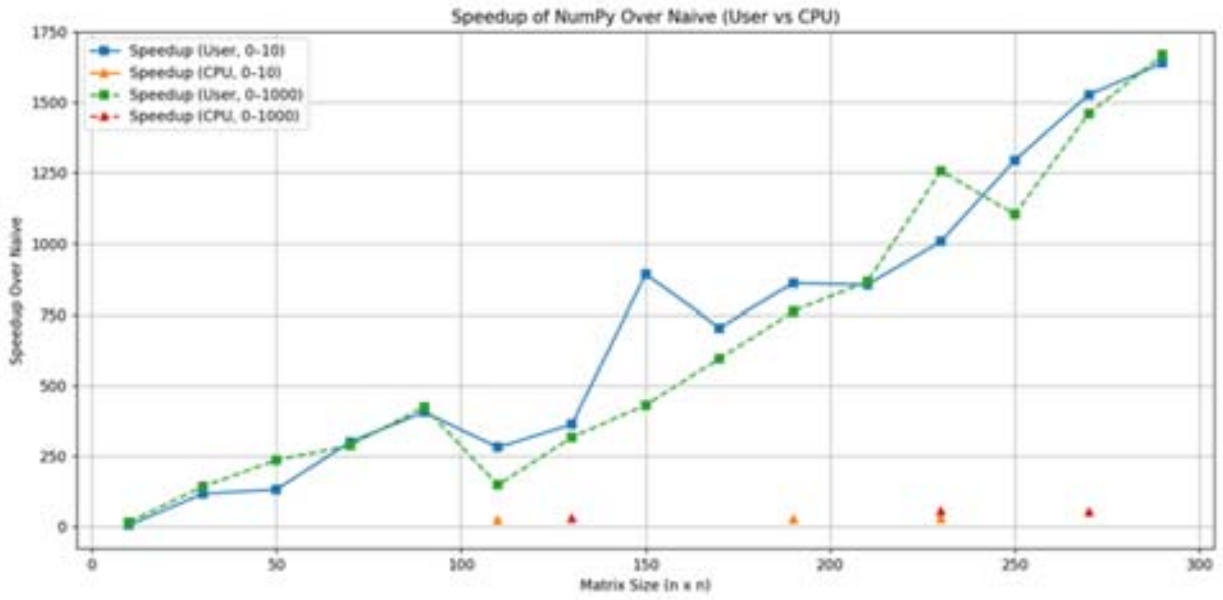

**Fig. 3:** Speedup of NumPy Over Naive Matrix Multiplication Using User and CPU Time for Value Ranges 0–10 and 0–1000.

Figure 3 compares the speedup of NumPy over Naive implementations. For both user and CPU time, NumPy delivers enormous performance gains, especially at larger matrix sizes. The user time speedup exceeds 1500x for large inputs, while the CPU time speedup shows less variability and reaches up to around 80x.

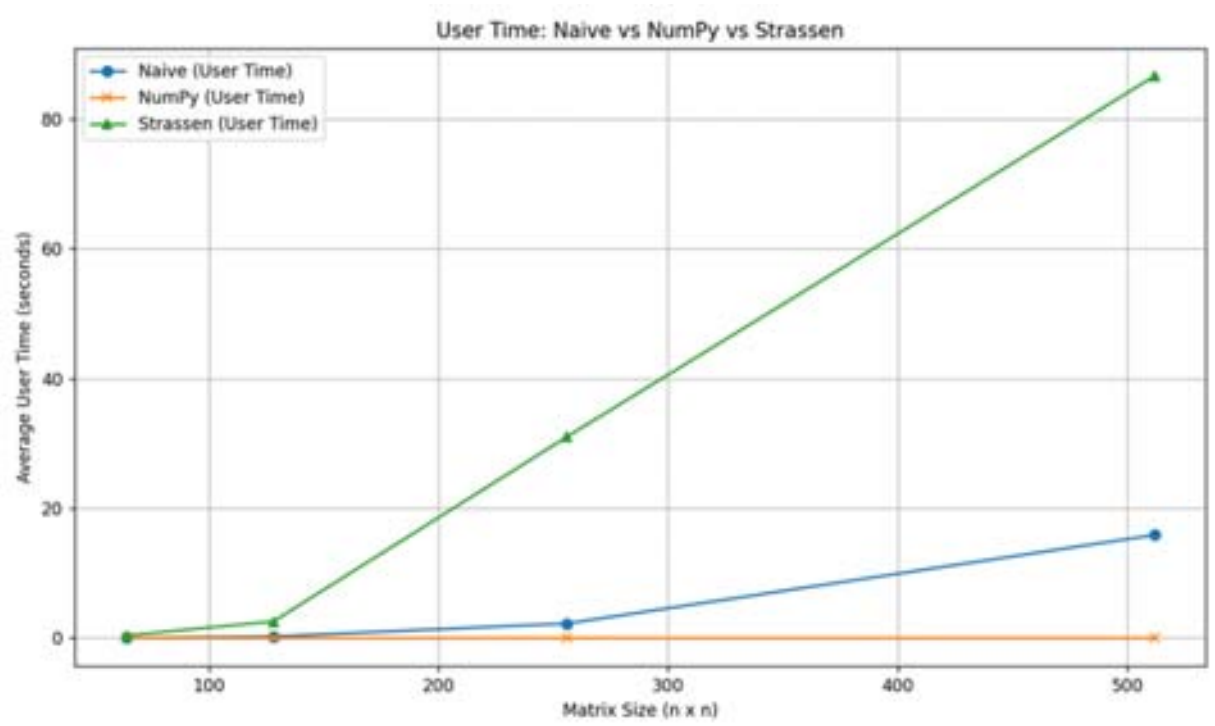

**Fig. 4:** Average User Time of Naive, NumPy, and Strassen Matrix Multiplication Algorithms.

Figure 4 presents the average user time across Naive, NumPy, and Strassen methods. While Strassen was theoretically expected to outperform Naive, the recursive overhead makes it significantly slower, especially for large matrices. NumPy remains the most efficient. The graph shows that NumPy has the lowest user time across all matrix sizes, indicating its high efficiency and optimization. Strassen's user time increases steeply for larger matrices, surpassing Naive's method due to its recursive overhead, despite its better theoretical complexity.

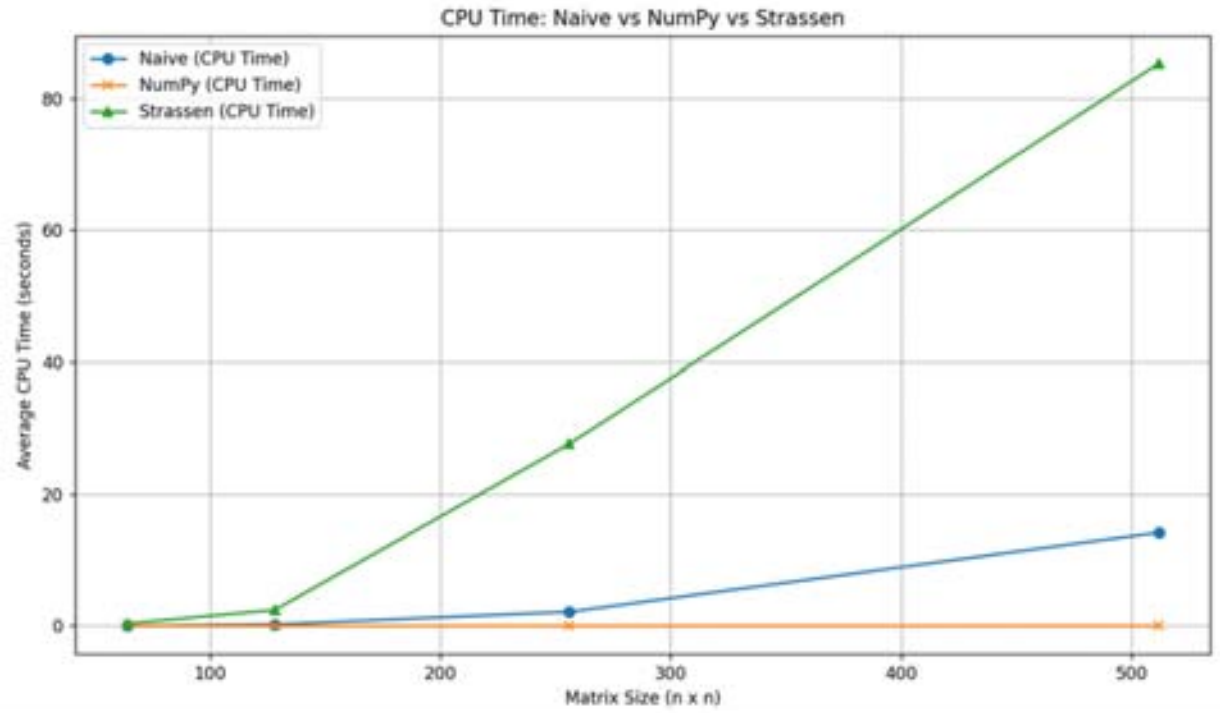

**Fig. 5:** Average CPU Time of Naive, NumPy, and Strassen Matrix Multiplication Algorithms.

Figure 5 shows the CPU time for the same methods. This graph shows that NumPy consistently maintains minimal CPU time regardless of matrix size, highlighting its optimization and low overhead. In contrast, Strassen's CPU time scales steeply, even exceeding Naive's at larger sizes, due to its recursive structure and memory overheads, despite its theoretical advantages.

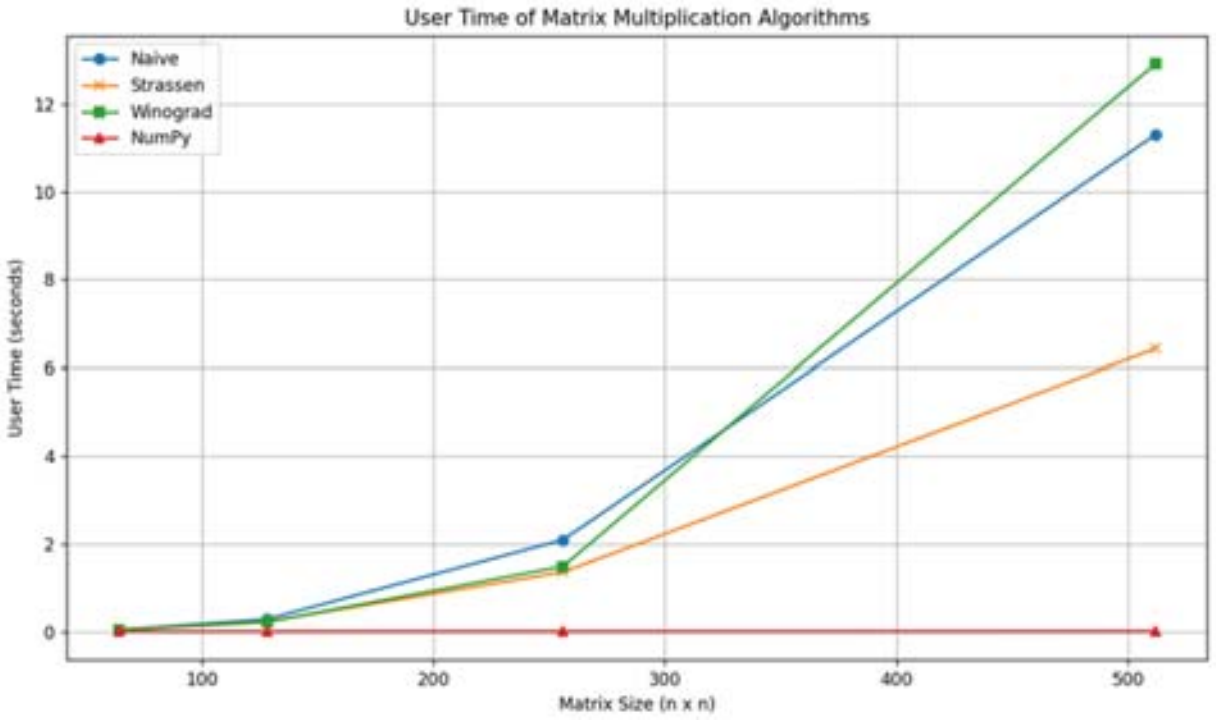

**Fig. 6:** Average User Time Comparison Among Naive, Strassen, Winograd, and NumPy Matrix Multiplication Algorithms.

Figure 6 expands the comparison to include the Winograd algorithm. Winograd performs slightly better than Naive for medium-sized matrices, but is overtaken by Naive as the size increases. Strassen shows some improvement over Naive at small sizes, but becomes increasingly inefficient at larger scales due to recursive overhead. NumPy consistently outperforms all other algorithms, maintaining the lowest user time across all matrix sizes. Overall, Winograd and Naive exhibit similar growth trends, with NumPy demonstrating superior efficiency.

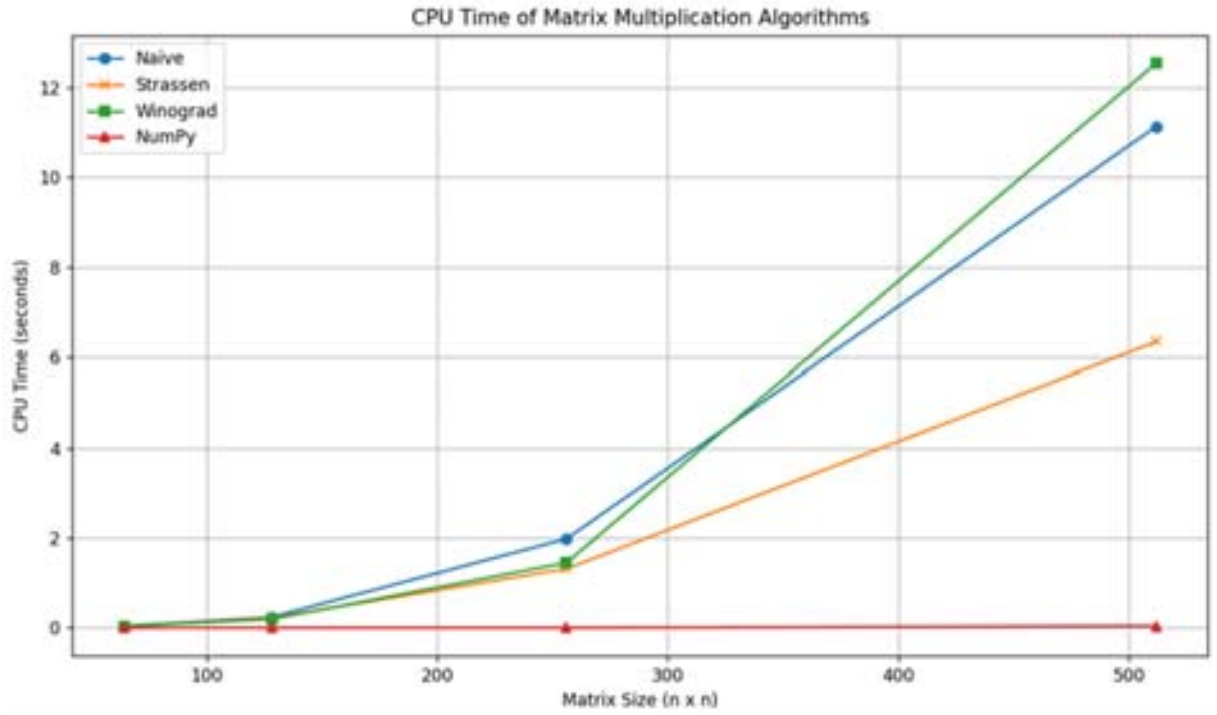

**Fig. 7:** Average CPU Time Comparison Among Naive, Strassen, Winograd, and NumPy Matrix Multiplication Algorithms.

Figure 7 shows that NumPy consistently achieves minimal CPU time, demonstrating its optimization over other methods. Naive and Winograd grow steeply in time with increasing matrix size, while Strassen initially performs better than Naive but becomes more costly at larger sizes. This suggests that recursive overhead in Strassen outweighs its divide-and-conquer benefits beyond moderate sizes.

### *B. Analysis of Graph Trends*

The plots shown in section A reveal several notable trends across different matrix multiplication methods and value ranges as stated below:

- **User vs. CPU Time**: Across all matrix sizes and value ranges, NumPy exhibits the lowest execution time both in user and CPU metrics, reflecting its optimized backend operations. Naive and Strassen implementations show a sharp increase in time with size, while Winograd offers moderate improvements.
- **Impact of Value Range**: Changing the value range (0–10 vs. 0–1000) has minimal effect on execution time, emphasizing that the performance is more sensitive to matrix size than value magnitude.
- **Speedup Observations**: Speedup graphs clearly indicate that NumPy outperforms naive by large margins (up to 1700× in some cases). Strassen and Winograd also show consistent speedup over naive but trail behind NumPy.
- **Scalability**: As matrix size grows, the execution time gap between naive and other algorithms widens, suggesting that optimization benefits compound with larger datasets.

These findings suggest that NumPy is ideal for production- scale matrix operations, while Strassen and Winograd offer educational insights into algorithmic acceleration and recursive computation strategies.

### *C. Hardware Acceleration*

In this section, we introduce the hardware acceleration based on a systolic array for a matrix multiplier. Matrix multiplication is a basic operation in various computations, such as digital signal processing, scientific simulation, and deep learning inference. For increasingly larger matrices, software-based solutions become inefficient since they inherently rely on a sequentially processed, complex calculation. The hardware option provides a more feasible approach, where the processes could be more efficiently parallelized, though with lower resolution.

In this section, we will establish the scope with respect to this research that is centered on the design and implementation of a parameterizable N×N systolic array using hardware description language. The proposed design entails the use of simple processing elements in a two-dimensional grid to enable scalable matrix multiplication. The proposed design aims to analyze the architectural performance, the time complexity, and the power dissipation of the resulting systolic array when the size varies.

The matrix multiplication process entails each corresponding element in the resulting matrix by taking their product. The computational complexity of matrix multiplication is given by $O(N^3)$. This operation is well-suited to be performed in parallel because each element in the output array can be computed independently, after the required data is gathered.

The systolic array architecture is comprised of a square array of processing elements that regularly perform multiply and accumulate operations and communicate with neighboring elements, as shown in the figure 8. Data is communicated horizontally and vertically through the array while the elements accomplish local multiply and accumulate operations. This approach leads to a highly scalable and simple hardware architecture with reduced global communication and ease of pipelining.

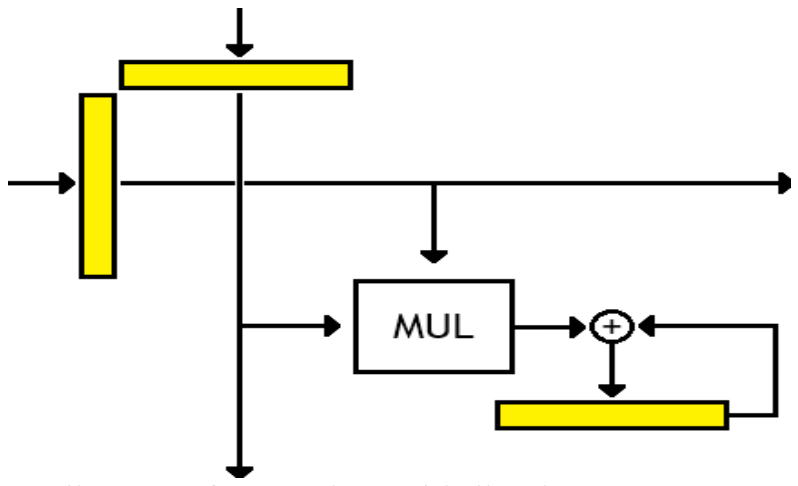

**Fig. 8.** Systolic array for matrix multiplication

The hardware architecture is based on a 2D systolic array of identical processing elements in an N x N grid. Data enters the array in a rhythmic procession, with matrix A elements with horizontal propagation, and matrix elements proactively propagating a vertical bit. Every processing element of the proposed model computes a local function and also transmits arguments to adjacent processing elements. This efficiently supports processing in parallel. Moreover, it does not require making any structural changes in their design to increase their size based on the requirement. The top-level flowchart implemented in this study is shown in Figure 9, and the flowchart of processing elements is shown in Figure 10.

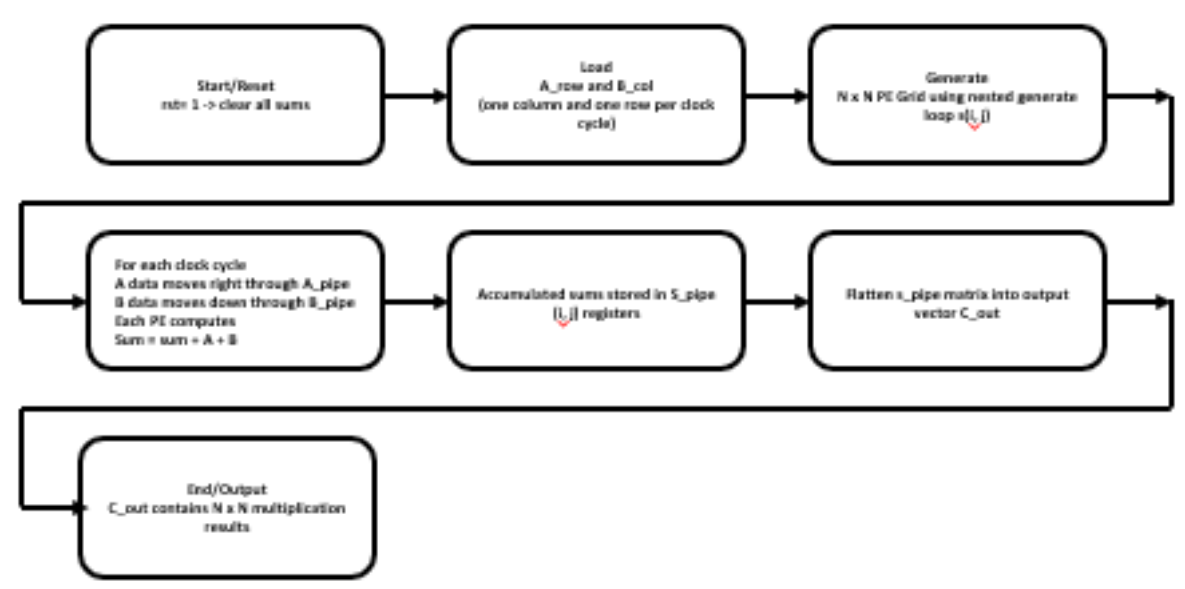

**Fig. 9:** Top-level flowchart of systolic array

Every processing element is intended to execute an unsigned multiply-accumulate operation synchronized by a global clock. One processing element takes input operands, calculates their product, and accumulates it into a sum stored in a register over several cycles. Then, it resets the sum synchronously to initiate processing of a computation. Input operands are passed directly to neighboring processing elements along the processing array, thus preserving a constant stream of input data flowing into the processing array. Control of the systolic array is handled by synchronized clocking and data fetching rather than control logic. The matrix data is streamed into the array one row and one column every clock cycle, allowing for the accumulation of partial sums of the data streaming across the grid. However, due to its pipeline structure, there exists an inherent latency based on the array size before producing valid results at regular cycles.

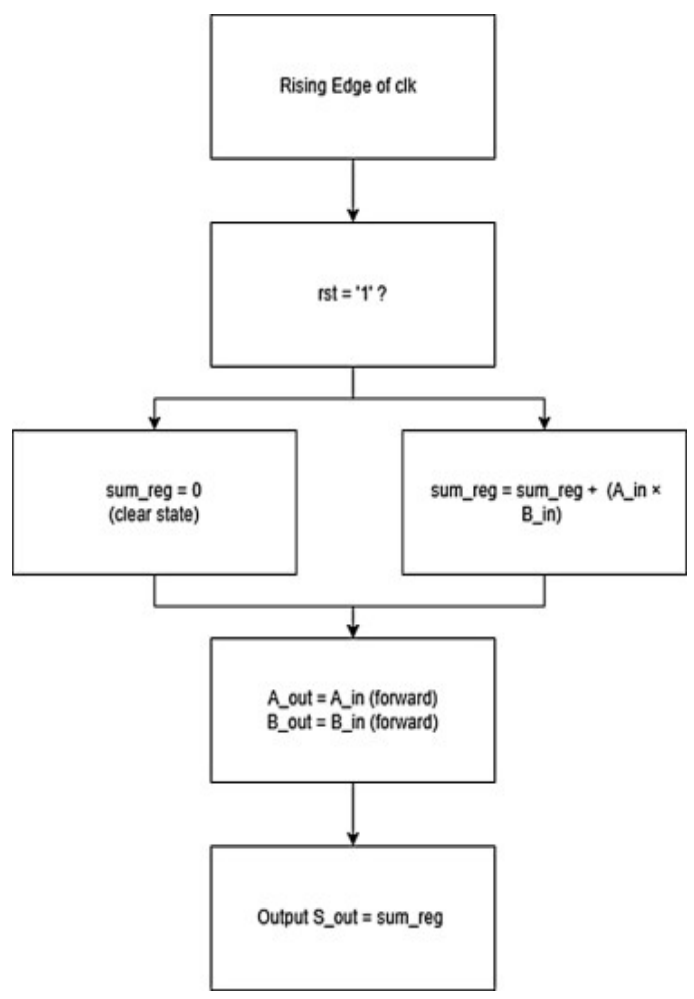

**Fig. 10** Flowchart of processing element

The systolic array design was carried out with VHDL. The design simulation was performed with a testbench. The testbench allowed the introduction of matrices to the design with structured stimulus. The results were checked to ensure proper functioning of the multiply and accumulate operations. The results of data processing were also checked with waveform analysis to ensure proper data alignment and the calculation of correct results. The design was synthesized after the verification stage. A snapshot of the simulated waveform is shown in Figure 11.

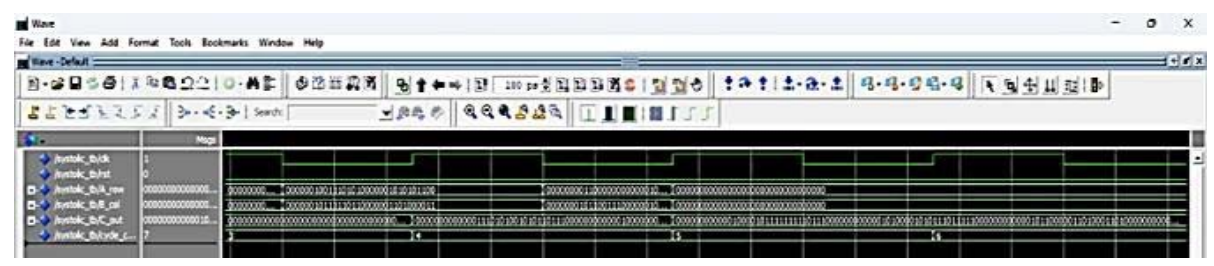

**Fig. 11:** Waveform for a 2x2 matrix multiplication

Performance evaluation was done based on analyzing the timing characteristics and power consumption for increasing array sizes. Maximum operating frequency and computational latency were derived using static timing analysis; power estimates were produced using post-synthesis power analysis tools. Switching activity information was included to enhance the accuracy of these estimates. Results over several array configurations are reported, thus giving the much-needed insight into the scalability impact on hardware performance and energy consumption.

**Table 1.** Time and power requirements for hardware acceleration

| Array size | Simulation Time (ns) | Hardware Time (ns) estimated | Power (mW) |
|---|---|---|---|
| 2x2 | 15 | 8 | 192.28 |
| 4x4 | 35 | 16 | 192.28 |
| 8x8 | 75 | 32 | 192.28 |
| 16x16 | 155 | 64 | 191.99 |
| 32x32 | 315 | 128 | 213.19 |
| 64x64 | 635 | 256 | 213.19 |
| 128x128 | 1275 | 512 | 205.26 |
| 256x256 | - | 1024 | 213.19 |
| 512x512 | - | 2048 | 195.66 |
| 1024x1024 | - | 4096 | 195.66 |

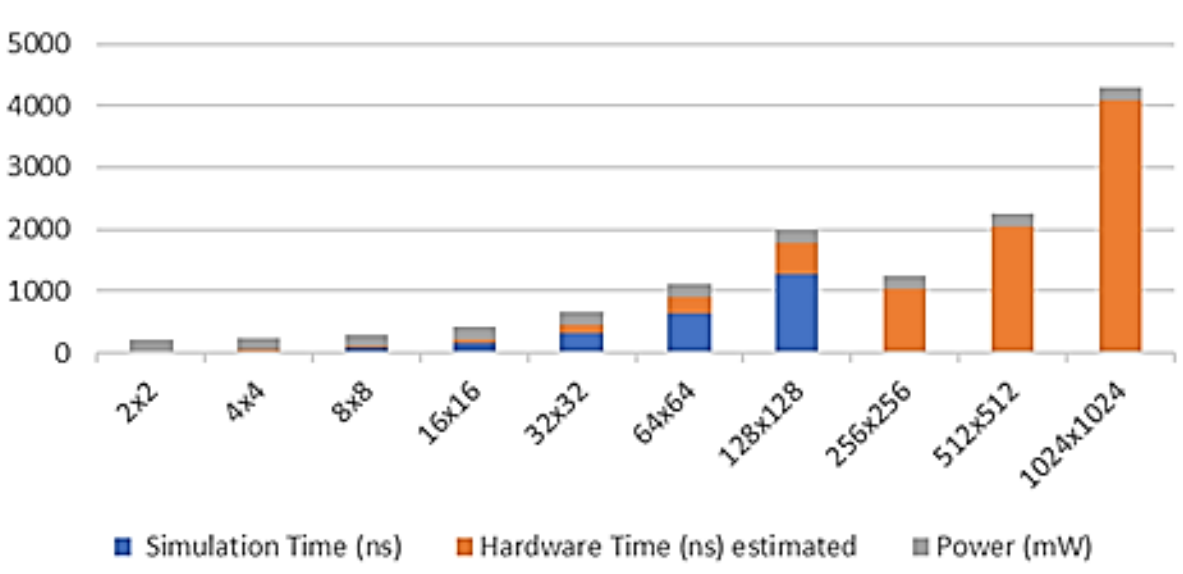

**Fig. 12.** Time and power breakdown for hardware acceleration

## IV. Discussion

As shown in Table 1 and Figure 12, compared to software-based matrix multiplication, the proposed systolic array–based hardware accelerator demonstrates significant advantages in execution time, scalability, and energy efficiency. Software simulations typically rely on sequential or limited parallel execution on general-purpose processors, leading to rapidly increasing computation time as the matrix size grows, particularly given the $O(N^3)$ complexity of matrix multiplication. In contrast, the proposed hardware architecture exploits massive spatial parallelism through a two-dimensional grid of processing elements, allowing multiple multiply-accumulate operations to be performed concurrently. As shown in Table 1, the estimated hardware execution time scales linearly with array size due to efficient pipelining, while software simulation times grow substantially and become impractical for large matrices. Furthermore, the power consumption remains relatively stable across increasing array dimensions, highlighting the energy efficiency of the hardware approach. This demonstrates that the systolic array architecture provides a more scalable and performance-efficient solution than software simulations, especially for large-scale matrix computations commonly encountered in signal processing and machine learning applications.

## IV. CONCLUSION AND FUTURE WORK

This paper presented a comparative evaluation of matrix multiplication algorithms across software and hardware platforms, highlighting how algorithmic design and hardware architecture jointly influence performance. The software analysis showed that while naïve implementations scale poorly, optimized libraries such as NumPy achieve substantial speedups through efficient use of cache hierarchies and vectorized execution. Although Strassen and Winograd reduce arithmetic complexity, their practical benefits on general-purpose CPUs are limited by recursion overhead and memory inefficiencies. In contrast, the proposed systolic array–based hardware accelerator demonstrates superior scalability, predictable latency growth, and improved energy efficiency by exploiting fine-grained parallelism and parallel MAC operations. However, this study is limited by its focus on square matrices, partially optimized software implementations, and reliance on estimated timing and power results for large hardware configurations. Future work will extend the evaluation to non-square matrices, mixed-precision arithmetic, and fully deployed FPGA implementations, as well as include broader comparisons with GPU-based accelerators to further explore software–hardware co-design trade-offs.